\documentclass[11pt,draftclsnofoot,onecolumn]{IEEEtran}

\usepackage{cite}
\usepackage[dvips]{color}
\usepackage[dvips,final]{epsfig}
\usepackage[dvips]{rotating}
\usepackage{bm}
\usepackage{amsmath}
\usepackage{amsfonts}
\usepackage{amssymb}
\usepackage{bm}
\usepackage{amssymb}
\usepackage{amsfonts}
\usepackage{amsmath}
\usepackage{psfrag}

\def\Tslot{T}
\def\bA{{\bf A}}
\def\bB{{\bf B}}

\def\bF{{\bf F}}

\def\bH{{\bf H}}
\def\bI{{\bf I}}

\def\bO{{\bf O}}

\def\bQ{{\bf Q}}
\def\bR{{\bf R}}
\def\bS{{\bf S}}

\def\bU{{\bf U}}

\def\bW{{\bf W}}

\def\bb{{\bf b}}
\def\bd{{\bf d}}
\def\bh{{\bf h}}

\def\bn{{\bf n}}

\def\bp{{\bf p}}

\def\br{{\bf r}}
\def\bs{{\bf s}}

\def\bu{{\bf u}}
\def\bv{{\bf v}}
\def\bw{{\bf w}}
\def\bx{{\bf x}}

\newcommand{\diag}[1]{\mathrm{diag}\left( #1 \right) }

\newcommand{\pdfchan}{\mathcal{CN}\left( 0,1\right)}

\def\trace{\mbox{tr} }

\newcommand{\expv}[1]{\mbox{E}\{ #1\}}
\newcommand{\expvv}[2]{\mbox{E}_{#2}\{ #1\}}
\newcommand{\myitemsep}{\setlength{\itemsep}{0mm}}
\newcommand{\RE}[1]{\mbox{Re}\left\lbrace  #1\right\rbrace}

\title{Cooperative Transmission for Wireless Relay Networks Using Limited Feedback}

\author{Javier M.~Paredes, Babak H.~Khalaj, and Alex B.~Gershman
\thanks{J.~M.~Paredes and
        A.~B.~Gershman are with the Communication
Systems Group, Darmstadt University of Technology, Merckstr.~25,
D-64283 Darmstadt, Germany; Phone: +49-6151-164970; Fax:
+49-6151-166095; e-mails: {\tt jparedes,
gershman@nt.tu-darmstadt.de}. B.~H.~Khalaj is with the Department
of Electrical Eng., Sharif University of Technology, Tehran, Iran;
e-mail: {\tt khalaj@sharif.ir}. The work of B.~H.~Khalaj was
supported by a Research Fellowship from the Alexander von Humboldt
Foundation. The work of A.~B.~Gershman was supported in part by
the European Research Council (ERC) under Advanced Investigator
Grant program and the German Research Foundation (DFG) under Grant
GE 1881/4-1.}}

\begin{document}

\maketitle

\begin{abstract}
To achieve the available performance gains in half-duplex wireless
relay networks, several cooperative schemes have been earlier
proposed using either distributed space-time coding or distributed
beamforming for the transmitter without and with channel state
information (CSI), respectively. However, these schemes typically
have rather high implementation and/or decoding complexities,
especially when the number of relays is high. In this paper, we
propose a simple low-rate feedback-based approach to achieve
maximum diversity with a low decoding and implementation
complexity. To further improve the performance of the proposed
scheme, the knowledge of the second-order channel statistics is
exploited to design {long-term power loading} through maximizing
the receiver signal-to-noise ratio (SNR) with appropriate
constraints. This maximization
 problem is approximated by a convex feasibility problem whose solution
is shown to be close to the optimal one in terms of the error
probability. Subsequently, to provide robustness against feedback
errors and further decrease the feedback rate, an extended version
of the distributed Alamouti code is proposed. It is also shown
that our scheme can be generalized to the differential
transmission case, where it can be applied to wireless relay
networks with no CSI available at the receiver.

\end{abstract}

\section{Introduction}
The performance of wireless communication systems can be severely
affected by channel fading. To combat fading, multi-antenna
systems are commonly used as in such systems, the existence of
independent paths between the transmitter and receiver can be used
to achieve a higher degree of diversity than in single-antenna
systems \cite{LS03, PaulrajBook, GerSid}. However, restrictions in
size and hardware costs can make the use of multi-antenna systems
impractical in wireless networks. Fortunately, similar independent
paths are also available in wireless networks with multiple
single-antenna nodes, where some nodes are used as relays that
help to convey the information through the network. Using such
relays between the transmitter and receiver nodes offers the
so-called \textit{cooperative diversity} and, hence, can be a good
alternative to using multiple antennas at the transmitter and/or
receiver. Several cooperation methods between network nodes have
been proposed based on different relaying strategies; see
\cite{LW03,SEA03a,SEA03b,GV05,NBK04,AES05,DH06} and references
therein.

Among different relaying approaches, techniques using the
amplify-and-forward relaying strategy are of especial practical
interest because they do not require any signal processing (such
as decoding or compression) at the relays.

The use of space-time codes (originally developed for
multi-antenna systems \cite{TSC98}, \cite{Ala98}) in a distributed
fashion has been proposed for relay networks in \cite{LW03} and
\cite{JH06} using the amplify-and-forward approach. In this
cooperative strategy, the source terminal first transmits the
information symbols to the relays. Then, the relays encode their
received signals and their conjugates in a linear fashion and
transmit them to the destination node. This can be viewed as
distributed space-time coding (DSTC). The DSTC techniques only
require the knowledge of the received signal powers at the relays
and can achieve the maximum diversity available in the network. In
\cite{JJ07}, orthogonal space-time block codes (OSTBCs)
\cite{Ala98}, \cite{TJC99} and quasi-orthogonal STBCs (QOSTBCs)
\cite{J01},\cite{SX04} have been used along with the DSTC strategy
of \cite{JH06}. Both these DSTC approaches have been shown to
offer maximum diversity, optimal diversity products, low maximum
likelihood (ML) decoding complexity, linear encoding of the
information symbols, and robustness against relay failures.
Unfortunately, for more than two relays, the maximum rate of
OSTBCs reduces \cite{Liang03}, the decoding delay increases, and
the linear ML decoding complexity is no longer achievable
\cite{JJ07}. Furthermore, QOSTBCs are only applicable to
particular cases with certain numbers of relays. In addition,
their decoding complexity is higher than that of OSTBCs.

In \cite{RR08}, four-group decodable DSTCs\footnote{For these
codes, it is possible to split the maximum likelihood decoding
problem into four independent subproblems.} for any number of
relays are proposed. Although this approach reduces the decoding
complexity as compared to the full ML decoder, its complexity is
still rather high, especially in the case of more than four
relays. To recover the simple symbol-by-symbol ML decoding
property of the distributed OSTBCs for more than two relays, the
use of the source-to-relay CSI at the relays has been proposed
\cite{RR07,JJ07}. However, as shown in \cite{JJ07}, this does not
improve the resulting diversity or coding gains.

Another promising approach to amplify-and-forward relaying in
wireless networks is distributed beamforming; see \cite{Lar03,
JJ08a, DCL08, HSGL08, DPP08, CGS} and refe\-rences therein. As
most of distributed beamforming techniques require the full
knowledge of the instantaneous CSI for both the source-to-relay
and relay-to-destination links and, moreover, require a feedback
link between the destination and relays, the complexities of these
techniques are rather high. To decrease the distributed
beamforming complexity, the use of quantized feedback for
selecting beamforming weights from a codebook has been proposed in
\cite{KJJ08}. However, the codebook design requires a costly
numerical optimization and the resulting codebook needs to be
transmitted to each relay every time when the channel statistics
or the transmitted powers change.

In this paper (see also \cite{PKG08} and \cite{PKG09}), we
consider a wireless network where each relay only needs to know
its average received signal power (which is a common assumption
for DSTCs) and further assume that one-bit feedback per relay is
available for every channel realization. The proposed scheme is
based on the ideas of partial phase combining (PPC)
\cite{HP98,MSO01} and the group coherent codes (GCCs) \cite{AG04}
originally introduced for traditional multiple-antenna systems. It
will be shown by means of an approximate symbol error rate (SER)
analysis that such a low-rate feedback is sufficient to achieve
maximum diversity with an additional coding (power) gain.
Furthermore, the proposed scheme will be shown to enjoy linear
decoding complexity and minimum decoding delay for any number of
relays. Although the best possible choice for the feedback bits
has to be found by a full search, we provide two much simpler
methods to judiciously choose these bits.

{It should be noted that several techniques related to the
proposed scheme have been developed in \cite{MHMB05, MHMB06,
JMR08} in the context of sensor networks. In these papers,
randomly generated relay beamformer phases are iteratively
selected based on a low-rate feedback. In particular, in
\cite{JMR08} the application of binary signaling to the approach
of \cite{MHMB05, MHMB06} has been considered. The scheme of
\cite{MHMB05, MHMB06, JMR08} requires multiple iterations to
converge, and the number of such iterations is in average
comparable to or larger than the number of sensor nodes. In
contrast to the approaches of \cite{MHMB05, MHMB06, JMR08}, the
proposed scheme will use a fixed (substantially lower) number of
feedback bits without any need for multiple iterations.}

Since the quality of channel links can vary for different relays,
we propose to use second-order channel statistics to find proper
``long-term'' {power loading coefficients}  for each relay. From
the performance viewpoint, these coefficients should be designed
by minimizing the error probability as it was proposed in
\cite{JJ08c} for a two-relay network using the distributed
Alamouti code. However, the approach of \cite{JJ08c} does not
provide any extension to the case of more than two relays. As an
alternative, we propose to use the general idea of \cite{HSGL08}
to obtain the {power loading coefficients} by maximizing the
average SNR subject to individual power constraints. However, in
contrast to \cite{HSGL08}, the loss in diversity is avoided by a
proper choice of the instantaneous feedback bits and by using
appropriate constraints on these coefficients. It is shown that
using semi-definite relaxation (SDR), the resulting SNR
maximization problem can be turned into a convex feasibility
problem which can be efficiently solved using interior point
methods. Simulations show that the resulting solution performs
very close to the direct (computationally prohibitive) approach
that minimizes the Chernoff bound on the error probability using
brute force optimization.

Using an extended version of the distributed Alamouti code, we
further refine the proposed technique to reduce the amount of
feedback without affecting the benefits of linear decoding
complexity and maximum diversity. In addition, the use of such an
extended distributed Alamouti code is shown to provide extra
robustness to erroneous feedback.

Finally, an extension of the proposed scheme for non-cohe\-rent
receivers using differential transmission is developed. It is
demonstrated that the proposed non-coherent scheme enjoys the same
advantages in performance, decoding complexity and delay as its
coherent counterpart.

The remainder of this paper is organized as follows. In Section
\ref{sysmodel}, the system model is developed. Section
\ref{sec:quasi} presents the proposed scheme. Its further
refinement using the extended distributed Alamouti code is
discussed in Section \ref{sec:eAlamouti}. The differential
transmission extension of the proposed techniques is developed in
Section \ref{sec:differential}. Computer simulations are presented
in Section \ref{sec:simulations} and conclusions are drawn in
Section \ref{sec:conclusions}.

\section{System Model} \label{sysmodel}
Let us consider a half-duplex wireless relay network with $R+2$
nodes where each node has a single antenna that can transmit or
receive signals. Among these $R+2$ nodes, one is the transmitter,
one is the receiver, and the remaining $R$ nodes are the relays.
It is assumed that the direct link between the transmitter and the
receiver can not be established and that the relay channels are
statistically independent. We consider the quasi-static flat
fading channel case with the block length $\Tslot$, and denote the
channel coefficient between the transmitter and the $i$th relay by
$f_i$. Correspondingly, the channel coefficient between the $i$th
relay and the receiver is denoted by $g_i$. We assume that $f_i$
and $g_i$ are independent random variables with the probability
density functions (pdf's) $\mathcal{CN}(\mu_{f_i},\sigma_{f_i}^2)$
and $\mathcal{CN}(\mu_{g_i},\sigma_{g_i}^2)$, respectively, where
$\mathcal{CN}( \cdot,\cdot)$ denotes the complex Gaussian pdf.

We assume that the transmitter does not have any CSI. However, we
consider a {limited} feedback link between the receiver and each
relay. This feedback link is used to transmit one bit for every
channel realization and can be also used to transmit long-term
{power loading coefficients} (one per relay) every time the
channel means or variances change significantly. The receiver may
or may not enjoy full CSI, depending on the transmission mode
(coherent or non-coherent) and the system is synchronized at the
symbol level.

At the transmitter side, $T$ symbols $\bs=[s_1,\ldots,s_T]^T$ are
drawn from an $M$-point constellation according to the information
bits to be sent. Here, $(\cdot)^T$ denotes the transpose. The
signal $\bs$ is normalized as $\expv{\bs^H \bs} = 1$, where
$(\cdot)^H$ and $\expv{\cdot}$ denote the Hermitian transpose and
the statistical expectation, respectively. The transmission is
carried out in two steps. In the first step, the transmitter sends
$\sqrt{P_0T}\bs$ from time $1$ to $T$, where $P_0$ is its average
transmitted power. The received signal at the $i$th relay is given
by
\begin{equation}
    \br_i=\sqrt{P_0T} f_i \bs + \bv_i
\end{equation}
where $\bv_i$ is the noise vector at the $i$th relay.
In the second step, the $i$th
relay sends the signal $\bd_i$ to the receiver from time $T+1$ to
$2T$. At the receiver, we have
\begin{equation}\label{eq:relay-rx}
    \bx=\sum_{i=1}^{R}g_i\bd_i + \bn
\end{equation}
where $\bx=[x_1,\ldots,x_T]^T$ is the received signal and $\bn$ is
the receiver noise vector. We assume that the entries of the noise
vectors $\bv_i$ and $\bn$ are i.i.d. random variables with the pdf
$\pdfchan$, that is, both these noises have variance $\sigma^2=1$.

The transmitted signal $\bd_i$ at each relay is assumed to be a
linear function of its received signal and its conjugate
\cite{JJ07}, that is,
\begin{eqnarray}
    \bd_i \!\!\!\! &=& \!\!\!\! \sqrt{\frac{P_i}{m_{f_i}P_0+1}}\, b_i
\theta_i(\bA_i\br_i+\bB_i\br^*_i)\nonumber\\
\!\!\!\!&=& \!\!\!\!\sqrt{\frac{P_0 P_i
    T}{m_{f_i}P_0+1}}\, b_i \theta_i(f_i\bA_i\bs+f^*_i\bB_i\bs^*)
 + \sqrt{\frac{P_i}{m_{f_i}P_0+1}}
b_i \theta_i(\bA_i\bv_i+\bB_i\bv^*_i) \label{eq:ti}
\end{eqnarray}
 where $m_{f_i} \triangleq
 \expv{|f_i|^2}= |\mu_{f_i}|^2+\sigma_{f_i}^2$,
  $b_i \in \{-1,\,1\}$ is a
coefficient selected based on the value of the one-bit feedback,
$\theta_i$ ($0 \leq \theta_i \leq 1$) is a real-valued long-term
{power loading coefficient} that is adjusted according to the
channel statistics (as it will be explained in
Section~\ref{sec:beam_weights}), $P_i$ is the maximum average
{power available} at the $i$th relay (while the actual
{transmitted} power is $P_i\theta_i^2\le P_i$), $(\cdot)^*$
denotes the complex conjugate and the $T\times T$ matrices $\bA_i$
and $\bB_i$
  are assumed to be either $\bA_i=\bO$ with
$\bB_i$ being unitary, or $\bB_i=\bO$ with $\bA_i$ being unitary.
Here, $\bO$ is the $T\times T$ matrix of zeros. {This assumption
implies that the statistics of the noise remains unchanged and
that the transmitted signal at each relay depends either on its
received signal or on the complex conjugate of this signal.}

Using this model, let us introduce the following notations:
\begin{eqnarray}
 \mbox{If } \bB_i=\bO &\!\!\mbox{then}\!\!& \tilde{\bA}_i=\bA_i,\,
 \tilde{f}_i=f_i,\,
\tilde{\bv}=\bv_i,\,\tilde{\bs}_i=\bs \\
\mbox{If } \bA_i=\bO &\!\!\mbox{then}\!\!&
\tilde{\bA}_i=\bB_i,\,\tilde{f}_i=f^*_i,\,
\tilde{\bv}=\bv^*_i,\,\tilde{\bs}_i=\bs^*.\label{eq:Ai_0}
\end{eqnarray}
Taking into account \eqref{eq:ti}-\eqref{eq:Ai_0}, the received
signal model \eqref{eq:relay-rx} can be written as
\begin{equation}\label{eq:in-out}
 \bx=\bS (\bp\odot\bh) + \bw
\end{equation}
where
\begin{equation}
\bS\triangleq[\tilde{\bA}_1\tilde{\bs}_1,\ldots,
\tilde{\bA}_R\tilde{\bs}_R] \nonumber
\end{equation}
 is the distributed space-time code matrix,
\begin{equation}
\bh = [h_1 ,\ldots, h_R]^T
\triangleq[\tilde{f}_1g_1,\ldots,\tilde{f}_Rg_R]^T
\nonumber
\end{equation} is the equivalent channel vector,
\begin{equation}\label{eq:noise}
 \bw = [w_1,\ldots,w_T]^T\triangleq\sum_{i=1}^{R}
\sqrt{\frac{P_i}{m_{f_i}P_0+1}}\, b_i \theta_i g_i \tilde{\bA}_i
\tilde{\bv}_i +\bn
\end{equation}
is the equivalent noise vector, \begin{equation} \label{eq:power_vector}
\bp \triangleq
\left[\sqrt{\frac{P_0P_1T}{m_{f_i}P_0+1}}\, b_1 \theta_1 , \ldots,
\sqrt{\frac{P_0P_RT}{m_{f_R}P_0+1}}\, b_R \theta_R\right]^T
\end{equation} and $\odot$ denotes the Schur-Hadamard
(element-wise) matrix product.

\section{The Proposed Cooperative Transmission Scheme}\label{sec:quasi}
In this section, we address the problem of selecting the
coefficients $b_i$ ($i=1,\ldots, R$) and the long-term {power
loading coefficients} $\theta_i$ ($i=1,\ldots, R$). We assume that
the value of $\bp \odot \bh$ is known at the receiver and there is
a perfect (error-free) low-rate feedback link between the receiver
and the relays. We will first introduce the transmission strategy
based on one-bit feedback per re\-lay to choose the coefficients
$b_i$ for every channel realization. It will be shown that this
transmission scheme achie\-ves maximum diversity. Subsequently, a
further improvement of this scheme will be considered using an
additional long-term real-valued {power loading coefficient} to
feed back from the receiver to each relay. These coefficients will
be computed using second-order channel statistics.

For the sake of simplicity, in this section we assume that $T=1$.
Hence, matrices $\bA_i$ and $\bB_i$ become scalars and it is
assumed that $\bA_i=1$ and $\bB_i=0$. Correspondingly, $\bx$,
$\bv_i$, $\bn$, $\bw$
 and $\bs$ become scalars as well. A more general case when $T>1$ (and when
 $\bA_i$ and $\bB_i$  are matrices rather than scalars) will be
 considered in Section \ref{sec:eAlamouti}.

\subsection{Using One-Bit Feedback Per Relay} \label{sec:one_bit_feed}
As in the case of one-bit feedback the long-term {power loading}
is not taken into account, all the relays transmit with the
maximum power $P_i$ (i.e., $\theta_i=1$ for $i=1,\ldots, R$). In
this particular case, the received signal model \eqref{eq:in-out}
reduces to
\begin{equation} \label{eq:in-out_bi}
 x={\bf 1}^T_R(\bp \odot \bh) s + w
\end{equation}
where ${\bf 1}_R$ is the $R\times 1$ column vector of ones. For
the sake of simplicity, the sub-indices in all scalar values are
hereafter omitted.

Using \eqref{eq:noise} and \eqref{eq:in-out_bi}, the noise power
can be expressed as
\begin{eqnarray}
 P_w \!\!\!\!&=&\!\!\!\! \expv{|n|^2}+\sum_{i=1}^{R}
 \frac{P_i}{m_{f_i}P_0+1}\expv{|v_i|^2} b_i^2 |g_i|^2 =1+\sum_{i=1}^{R} \frac{P_i}{m_{f_i}P_0+1}
|g_i|^2.\label{eq:noise_pow1}
\end{eqnarray}
{From} \eqref{eq:noise_pow1} it is clear that the choice of $b_i$
does not affect the noise power. Using \eqref{eq:in-out_bi}, the
signal power can be obtained as
\begin{eqnarray}
P_s \!\!\!\!&=&\!\!\!\! | {\bf 1}^T_R(\bp \odot \bh)|^2
 \expv{|s|^2} \nonumber\\
\!\!\!\!&=&\!\!\!\! \left | \sum_{i=1}^{R} \sqrt{\frac{P_0P_i}{m_{f_i}P_0+1}}
b_i f_i g_i \right|^2 = \underbrace{\sum_{i=1}^{R} \rho_{i,i}
|f_i g_i|^2}_{\gamma} +
\underbrace{\sum_{\substack{ i,j = 1\\ i \neq j }}^{R} \rho_{i,j} b_i b_j
\RE{f_i g_i f_j^* g_j^*}}_{\beta} \label{eq:Ps}
\end{eqnarray}
where \begin{eqnarray} \rho_{i,j}\triangleq\sqrt{\frac{T P_0 P_i
}{m_{f_i}P_0+1}}\sqrt{ \frac{T P_0 P_j}{m_{f_j}P_0+1}}\label{rho}
\end{eqnarray}
for $i,j=1,\ldots,R$, and $\RE{\cdot}$ denotes the
real part operation. In general, $\beta$ can take
 negative values. Clearly, such negative values of $\beta$ will reduce the received SNR and
affect the achieved diversity. Our key idea here is to use the
coefficients $b_i$ to ensure that $\beta$ is always non-negative.
It can be proved using the same approach as presented in
\cite{AG04} that using values of $b_i \in \{-1,1\}$ is sufficient
to guarantee $\beta\geq 0$. This results in a scheme with the
diversity order proportional to $R$, as stated in the following
proposition.

\textit{Proposition 1:} If $\beta \geq 0$, then the average symbol
error probability ($\overline{\mathrm{SER}}$) for
\eqref{eq:in-out_bi} can be upper bounded by
\begin{equation}\label{eq:prop1}
 \overline{\mathrm{SER}} \leq
\kappa P^{-R(1-\frac{\log{\log{P}}}{\log{P}})}
\end{equation}
for large $R$ and large $P$, where $P$ is the total power in the
network and $\kappa$ is a constant.

\textit{Proof:} See Appendix.

It follows from Proposition 1 that the achievable diversity order
of the proposed scheme is $R$.

Since positive values of $\beta$ will provide an additional signal
power gain, the optimal values of $b_i$ ($i=1,\ldots, R$) can be
obtained through maximizing $\beta$. This is an integer
maximization problem that requires a full search
 over all possible values of $b_i$. Clearly, if
the number of relays is large, then such a full search procedure
can be impractical. To reduce the complexity, we propose a
near-optimal solution based on SDR, that we denote hereafter as
\textit{Algorithm 1}.

Note that, according to (\ref{eq:Ps}), the choice of $b_i$ does
not affect the value of $\gamma$. Therefore, to maximize $P_s$, it
is sufficient to maximize $\beta$ in (\ref{eq:Ps}). Let us express
$P_s$ in a more convenient form by extending the notation for
$\rho_{i,j}$ in (\ref{rho}) with $$\rho_{i,0} \!\triangleq
\!\!\sqrt{\frac{P_i P_0 T}{m_{f_i}P_0+1}}$$ and denoting
\begin{equation}\label{eq:h_w_power}
    \bar{\bh}\triangleq \left [
\rho_{1,0} f_1g_1, \ldots,\rho_{R,0} f_Rg_R
      \right ]^T.
\end{equation}
Using  \eqref{eq:h_w_power}, the signal power \eqref{eq:Ps} can be
expressed as
\begin{equation}
 P_s=  |\bar{\bh}^H\bb|^2
 \nonumber
\end{equation}
where $\bb \triangleq [b_1, \ldots , b_R]^T$. Defining
$\bar{\bQ}\triangleq \bar{\bh}\bar{\bh}^H$, we can write the
optimization problem as
\begin{equation}\label{eq:op_Ps}
    \max_{\bb \in \{-1,1\}^R} \bb^T \bar{\bQ} \bb.
\end{equation}
As $\bb^T \bar{\bQ} \bb=\trace(\bb\bb^T\bar{\bQ})$, the
optimization problem in \eqref{eq:op_Ps} can be rewritten as
\begin{eqnarray}\label{eq:op_Ps2}
    \max_{{\bB}} \,\trace({\bB}\bar{\bQ})\quad
\mbox{s.t. } &\!\!\!\!\!\!& \mbox{ rank}\{ {\bB}\} = 1, \ \ {\bB} \succeq 0, \ \  {[{\bB}]}_{ii}=1, \quad i=1,\,\ldots,\,R
\end{eqnarray}
where ${\bB} \triangleq \bb\bb^T$, $\bb \in \mathbb{R}^R$,
$\trace(\cdot)$ stands for the trace of a matrix, $\mathbb{R}$
denotes the set of real numbers, and $[{\bB}]_{ii}$ denotes the
$i$th diagonal element of ${\bB}$. Problems similar to
\eqref{eq:op_Ps2} arise in the context of ML detection. Solutions
close to the optimal one can be efficiently found using the SDR
approach \cite{MDWLC02}, whose essence is to omit the rank-one
constraint $\mbox{ rank}\{ {\bB}\} = 1$ in (\ref{eq:op_Ps2}) and,
therefore, approximate the latter non-convex problem by a convex
problem
\begin{eqnarray}\label{eq:op_Ps2mod}
    \max_{{\bB}} \,\trace({\bB}\bar{\bQ})\quad
\mbox{s.t. } &\!\!\!\!\!\!\!\!& {\bB} \succeq 0,\
{[{\bB}]}_{ii}=1, \ i=1,\,\ldots,\,R.\ \ \
\end{eqnarray}
{Note that this problem can be efficiently solved using interior
point techniques \cite{MDWLC02, BV04, SDL06}. Generally, the
resulting solution for $\bB$ is not guaranteed to be rank-one. If
it is rank-one, then its principal eigenvector is the optimal
solution to (\ref{eq:op_Ps}). Otherwise, a proper approximate
solution for $\bm{\theta}$ can be recovered from $\bB$ using
randomization techniques, see \cite{MDWLC02} and \cite{SDL06} for
more detail.}

Thus, our SDR-based approach can be summarized as follows.

 \underline{\textit{Algorithm 1}}
\begin{itemize}
\item[1.] At the receiver, find the solution to \eqref{eq:op_Ps2mod}
using the approach of \cite{MDWLC02}.
\item[2.] Send the so-obtained $b_i$ from the receiver to the $i$th relay node
for each $i=1,\ldots, R$ using one-bit per relay feedback.
\end{itemize}

As it will be shown throughout our simulations, the use of the SDR
approach results in a performance that is very close to that of
the full search-based approach. The complexity of the SDR approach
is much lower than that of the full search; see \cite{MDWLC02} for
details.

To further reduce the complexity, let us discuss a simpler
algorithm to obtain acceptable values of $b_i$ that can be
formulated using the general idea of \cite{AG04}. The essence of
this algorithm is to use a greedy selection of the values of $b_i$
in a consecutive way. This algorithm can be summarized as the
following sequence of steps:

\underline{\textit{Algorithm 2}}
\begin{itemize}
\myitemsep
\item[1.] Set $b_1=1$ and $\tau_1=h_1$.
\item[2.] For $i=2,\,\ldots,\,R$, compute
\begin{eqnarray*}
b_i&\!\!\!=\!\!\!& \mbox{sign}(\RE{h_i^*\tau_{i-1}}), \ \ \
\tau_i=  \tau_{i-1}+ b_i h_i \end{eqnarray*}
where
$\mbox{sign}(\cdot)$ is the sign function.
\item[3.] Send the so-obtained $b_i$ from the receiver to the $i$th relay node
for each $i=1,\ldots, R$ using one-bit per relay feedback.
\end{itemize}

Note that Algorithm 2 does not result in the optimal values of
$b_i$, $i=1,\ldots, R$. However, Algorithm 2 is computationally
much simpler than Algorithm 1. Hence, these two alternative
techniques are expected to provide different
performance-to-complexity tradeoffs.

\subsection{Choosing Long-Term {Power Loading}}\label{sec:beam_weights}
So far, we have not considered the use of {power loading},
$\theta_i$, for each relay. In practical scenarios, relays are
distributed randomly in an area between the transmitter and the
receiver. As a result, the power loss characteristics in the
source-to-relay and relay-to-destination links are different for
each relay. Furthermore, different relays may have different
transmitted power constraints. Therefore, in such scenarios, some
{power loading} strategy should be employed to take into account
such differences in channel quality and/or constraints on the
relay transmitted power.

{From} the performance viewpoint, the optimal power loading should
be designed by minimizing the error probability as proposed in
\cite{JJ08c} for a two-relay network using the distributed
Alamouti code. However, the approach of \cite{JJ08c} does not
provide any extension to the case of more than two relays. As an
alternative to the error probability criterion, we propose to use
the general idea of \cite{HSGL08} to obtain the power loading
coefficients by maximizing the average SNR subject to individual
power constraints.

In what follows, the maximization of the average received SNR is
used as a criterion to design the {power loading} coefficients
$\theta_i$. Note that a related strategy to choose the beamforming
weights was also used in \cite{HSGL08}. However, we will show that
in contrast to \cite{HSGL08}, full diversity can be achieved in
our case by using the optimal feedback values $b_i$ along with the
coefficients $\theta_i$.

First, let us evaluate the average signal power, that is
\begin{equation}\label{eq:new_exp_Ps}
\expv{P_s}=\expv{\tilde{\gamma}}+\expv{\tilde{\beta}}
\end{equation}
where
\begin{eqnarray}
 \tilde{\gamma} & \!\!\triangleq\!\! & \sum_{i=1}^{R} \rho_{i,i}\theta_i^2 |f_i g_i|^2\label{eq:gamma_new}\\
 \tilde{\beta} & \!\!\triangleq\!\! & \sum_{\substack{ i,j = 1\\ i \neq j }}^{R} \rho_{i,j} \theta_i  \theta_j b_i b_j
\RE{f_i g_i f_j^* g_j^*}.\label{eq:beta_new}
\end{eqnarray}

Note that in \eqref{eq:new_exp_Ps}, the analytical evaluation of
$\expv{\tilde{\beta}}$ is very difficult due to the dependence of
$b_i$ ($i=1,\ldots, R$) on the instantaneous channel values.
Therefore, using \eqref{eq:gamma_new}-\eqref{eq:beta_new} and
assuming that the optimal $b_i (i=1,\ldots,R)$ are selected, we
propose to approximate \eqref{eq:new_exp_Ps} as
\begin{eqnarray}
\expv{P_s} \!\!\!\!&\approx&\!\!\!\!  \sum_{i,j=1}^{R} \rho_{i,j}
\theta_i \theta_j
 |\RE{\expv{f_i g_i f_j^* g_j^*}}|.
\label{eq:exp_Ps}
\end{eqnarray}

\begin{figure}[!b]
\begin{center}
  \includegraphics[width=7.9cm]{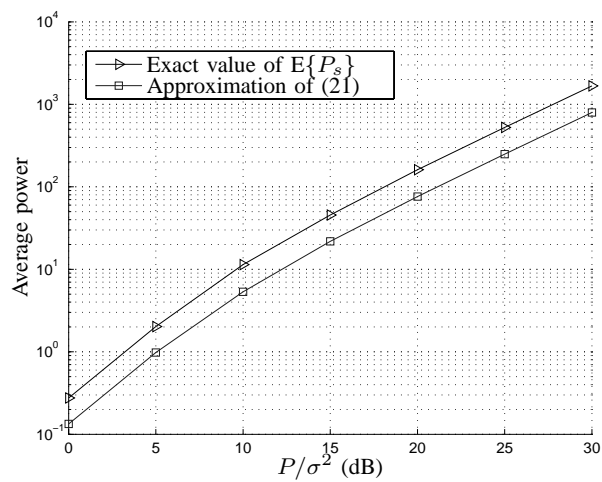}
\end{center}
\caption{Comparison of the approximation \eqref{eq:exp_Ps} and the
exact value of
  $\mbox{E}\{P_s\}$.}
  \label{fig:ComparisonApproxVsReal}
\end{figure}

The quality of this approximation is illustrated in
Fig.~\ref{fig:ComparisonApproxVsReal} where the exact value of
$\expv{P_s}$ and its approximation (\ref{eq:exp_Ps}) are plotted
versus $P$ normalized by the noise variance $\sigma^2$. In this
figure, it is assumed that $R=4$ and $\theta_i=1$ ($i=1,\ldots ,
R$). All the channels are assumed to be complex circular Gaussian
random variables with zero-mean and unit variance.

Another important question when using this approximation is how
close the values of achieved average SNR obtained from the
approximation in (\ref{eq:exp_Ps}) and from the exact value of
$\expv{P_s}$ are. This question was investigated by means of
extensive Monte-Carlo simulations that, for the sake of brevity,
are not presented in all detail. These simulations have involved
different channel scenarios, randomly generated channel
coefficients for each particular scenario, and different numbers
of relays lying in the interval $R=2,\ldots,7$. The optimal
coefficients $\theta_i$ have been obtained by brute force
optimization of $\expv{P_s}$ and their approximate values have
been found by optimizing (\ref{eq:exp_Ps}). Then, the achieved
average SNRs were compared for two so-obtained sets of optimized
power loading coefficients.

The results of this comparison have verified that the difference
between the exact optimal SNR (computed numerically via brute
force optimization of $\expv{P_s}$) and its approximation computed
via (\ref{eq:exp_Ps}) is, in average, less than $3\%$. This
implies that the approximation (\ref{eq:exp_Ps}) is worth using
for maximizing the average SNR by power loading.

Using the statistical independence of all source-to-relay and
relay-to-destination channels, we can now estimate the expected
value in \eqref{eq:exp_Ps} as
\begin{eqnarray}
 \expv{f_i g_i f_j^* g_j^*} = \expv{f_i f_j^*}\expv{g_i g_j^*}=
 ( \mu_{f_i} \mu_{f_j}^* +
\delta_{ij}\sigma^2_{f_i}) (
\mu_{g_i}\mu_{g_j}^*+\delta_{ij}\sigma^2_{g_i})\,\,
\label{eq:exp_calc}
\end{eqnarray}
where $\delta_{ij}$ is the Kronecker
delta.

Let us define the real-valued matrix $\bQ$ with the $(i,j)$ entry as
 \begin{equation} \label{eq:Qmat}
[\bQ]_{i,j}=\rho_{i,j} \left|\RE{ ( \mu_{f_i} \mu_{f_j}^* +
\delta_{ij}\sigma^2_{f_i}) (
\mu_{g_i}\mu_{g_j}^*+\delta_{ij}\sigma^2_{g_i})}\right|
\end{equation}
for $i,j=1,\,\ldots,\,R$.
 Using (\ref{eq:exp_calc}) and \eqref{eq:Qmat}, equation \eqref{eq:exp_Ps} can
be written as
\begin{equation}\label{eq:e_Ps}
 \expv{P_s} \approx \bm{\theta}^T\bQ \bm{\theta}
\end{equation}
where $\bm{\theta}\triangleq [\theta_1, \ldots, \theta_R]^T$.
 Using the fact that the noise waveforms and the channel
coefficients are statistically independent, the noise power can be
expressed as
\begin{eqnarray}
\expv{P_w} \!\!\!\!&=&\!\!\!\! \expv{|n|^2}+\sum_{i=1}^{R}
\frac{\theta_i^2 P_i}{m_{f_i}P_0+1}\,\expv{|v_i|^2} \expv{|g_i|^2} = 1+\sum_{i=1}^{R} \frac{\theta_i^2 P_i
m_{g_i}}{m_{f_i}P_0+1} \nonumber
\end{eqnarray}
and further rewritten as
\begin{equation}\label{eq:e_Pw}
 \expv{P_w} = \bm{\theta}^T\bW \bm{\theta}+1
\end{equation}
where $$ \bW \triangleq \diag{\frac{m_{g_1} P_1 }{m_{f_1}P_0+1}
,\, \ldots, \, \frac{m_{g_R} P_R}{m_{f_R}P_0+1}}$$ and
$\diag{\cdot}$ denotes a diagonal matrix. Using \eqref{eq:e_Ps}
and \eqref{eq:e_Pw}, the maximization of the receiver SNR over
$\bm{\theta}$ can be approximately written as
\begin{eqnarray}\label{eq:op_SNR}
    \max_{\bm{\theta}} \,\frac{\bm{\theta}^T\bQ \bm{\theta}}
{\bm{\theta}^T\bW \bm{\theta}+1}\quad\quad \mbox{s.t. }\quad
\theta_i^2\leq 1, \ \  i=1,\,\ldots,\,R
\end{eqnarray}
where instead of the signal power we use its approximation given
by (\ref{eq:exp_Ps}).

 If the aggregate power constraint
($\bm{\theta}^T\bm{\theta}=R$) is used instead of the
individual relay power constraints in \eqref{eq:op_SNR}, the
resulting problem becomes
\begin{eqnarray}\label{eq:op_SNRmod}
    \max_{\bm{\theta}} \,\frac{\bm{\theta}^T\bQ \bm{\theta}}
{\bm{\theta}^T(\bW + (1/R)\bI_R)) \bm{\theta}}\quad\quad
\mbox{s.t. }\quad \bm{\theta}^T\bm{\theta}=R.
\end{eqnarray}
Solving (\ref{eq:op_SNRmod}) amounts to the unconstrained
optimization of the objective function in (\ref{eq:op_SNRmod})
(that boils down to solving a generalized eigenvector problem)
followed by rescaling the so-obtained vector $\bm{\theta}$ to
satisfy the constraint $\bm{\theta}^T\bm{\theta}=R$.

In what follows, we consider a more practical case of individual
power constraints rather than the aggregate power constraint.

As mentioned above, the design of {power loading coefficients}
 by maximizing the average SNR does not take into
account the diversity aspect of the problem. In fact, maximizing
the average SNR can result in a solution with a poor performance
in terms of the error probability. This can particularly be the
case if some of the resulting values of $\theta_i$ are small, so
that the diversity order suffers. {Indeed, if $\theta_i$ is close
to zero at the $i$th relay, then this is equivalent to switching
off the $i$th relay for all the transmissions within the time
interval where the current value of $\theta_i$ is used. According
to Proposition 1, this will reduce the diversity order.}

To prevent such a loss in diversity, an additional constraint
$\theta_i^2 \geq \bar{\theta}_i^2$ can be used where
$\bar{\theta}_i^2$ is a preselected minimum power loading value
that establishes a tradeoff between the diversity and {power
loading} performance. If $\bar{\theta}_i$ is chosen too large,
then the interval for $\theta_i$ will be smaller, and this may
prevent the scheme from achieving any significant improvement in
the {performance due to power loading}. Reversely, if
$\bar{\theta}_i$ is chosen too small, a substantial diversity loss
can occur.

Defining $\bm{\Theta}\triangleq \bm{\theta}\bm{\theta}^T$, we can
rewrite \eqref{eq:op_SNR} as
\begin{eqnarray}\label{eq:op_SNR1}
    \max_{\bm{\Theta}} \,\frac{\trace(\bQ\bm{\Theta})}{\trace(\bW \bm{\Theta})+1}\quad
&\!\!\mbox{s.t.}\!\!& \bar{\theta}_i^2\leq[\bm{\Theta}]_{ii} \leq 1,
\ \  i=1,\,\ldots,\,R \quad\nonumber\\
&\!\!\!\!&\mathrm{rank}\{\bm{\Theta}\} = 1, \quad
\bm{\Theta}\succeq 0 \quad
\end{eqnarray}
where $\bm{\Theta}\succeq 0$ means that $\bm{\Theta}$ is positive
semi-definite. Introducing the auxiliary variable $t$,
\eqref{eq:op_SNR1} can be written as
\begin{eqnarray}
    \max_{\bm{\Theta},t} \,t \quad
\mbox{s.t.}&\!\!& \trace(\bm{\Theta}(\bQ-t\bW)) \geq t\quad\quad\quad\nonumber\\
 &\!\!& \bar{\theta}_i^2\leq{[\bm{\Theta}]}_{ii} \leq 1, \ \  i=1,\,\ldots,\,R
 \quad \label{eq:op_SNR2}\\
&\!\!& \mathrm{rank}\{\bm{\Theta}\} = 1, \quad \bm{\Theta}\succeq
0 . \nonumber
\end{eqnarray}
As the rank constraint in \eqref{eq:op_SNR2} is non-convex, this
optimization problem can not be solved efficiently. Using the SDR
approach (i.e., ignoring the constraint
$\mathrm{rank}\{\bm{\Theta}\} = 1$ in (\ref{eq:op_SNR2})), a
quasi-convex optimization problem can be obtained from
\eqref{eq:op_SNR2} that can be directly solved using the bisection
technique \cite{BV04,HSGL08}. Based on the latter technique, the
optimal value $t_{\mathrm{opt}}$ is found in the interval
$[t_{\mathrm{low}},t_{\mathrm{up}}]$, where $t_{\mathrm{low}}$ is
a feasible value and therefore, $t_{\mathrm{opt}}\geq
t_{\mathrm{low}}$, and $t_{\mathrm{up}}$ is not a feasible value
and therefore, $t_{\mathrm{opt}}\leq t_{\mathrm{up}}$. The
algorithm solves the feasibility problem
\begin{eqnarray}\label{eq:op_SNR3}
    \mathrm{find}\  \bm{\Theta}\quad\quad
\mbox{s.t. }& & \trace(\bm{\Theta}(\bQ-t\bW)) \geq t,\quad \bm{\Theta}\succeq 0   \nonumber\\
& & \bar{\theta}_i^2\leq{[\bm{\Theta}]}_{ii} \leq 1, \ \  i=1,\,\ldots,\,R
\quad\quad\,\,
\end{eqnarray}
at the midpoint of the interval,
$t=(t_{\mathrm{low}}+t_{\mathrm{up}})/2$. If it is feasible,
$t_{\mathrm{low}}$ is updated as $t_{\mathrm{low}}=t$. If it is
not feasible, $t_{\mathrm{up}}$ is updated as $t_{\mathrm{up}}=t$.
Then, the algorithm continues to solve the feasibility problem
with the new interval until
 $t_{\mathrm{up}}-t_{\mathrm{low}}<\epsilon$, where $\epsilon$
is a parameter denoting the acceptable tolerance of the solution.
The optimal matrix ${\bm{\Theta}_\mathrm{opt}}$ is selected as
$\bm{\Theta}$ for the last feasible $t$, (i.e.,
$t=t_{\mathrm{low}}$ in the last step). If the matrix
${\bm{\Theta}_\mathrm{opt}}$ is rank-one, then its principal
eigenvector is the optimal solution to \eqref{eq:op_SNR}. If
${\bm{\Theta}_\mathrm{opt}}$ is not rank-one, then a proper
approximate solution for $\bm{\theta}$ can be obtained using
randomization techniques \cite{SDL06}.

\section{Extended Distributed Alamouti Code}\label{sec:eAlamouti}

The scheme developed in the previous section applies to the case
of $T=1$. In what follows, we extend it to the case of $T=2$ by
developing an approach based on the distributed Alamouti code to
reduce the total feedback rate. Using computer simulations, the
latter scheme will be shown to provide robustness against feedback
errors. Such improvements in the feedback rate and robustness are,
however, achieved at the price of an increased decoding delay and
a moderate performance loss as compared to the case of $T=1$.

Let us consider the case of an even number of relays\footnote{The
case of an odd number of relays can be addressed in the same way
and, therefore, is omitted below.}, i.e., let $R=2K$ where $K$ is
some positive integer. The distributed Alamouti code is used by
relay pairs. Let each $k$th relay pair receive a low-rate feedback
to select the binary coefficient $b_k \in \{-1,\,1\}$ and the
real-valued {power loading} coefficient $\theta_k \in [0,\,1]$.
Since the same $b_k$ and $\theta_k$ should be used by the two
relays of the $k$th relay pair, the receiver can broadcast them to
both these relays, thereby reducing the feedback rate almost by
half.

The relays use the basic distributed Alamouti code matrices \cite{JJ07} to
 form the signal transmitted by each relay pair as:
$\tilde{\bA}_{2k-1}=\bI_2$ (with $\bB_{2k-1}=\bO$) and
$
 \tilde{\bA}_{2k}=\left[ \begin{array}{rr}
 0 & -1\\
 1 & 0
\end{array}
\right]$ with $\bA_{2k}=\bO$, where $\bI_2$ is the $2\times 2$
identity matrix. Using \eqref{eq:in-out}-\eqref{eq:power_vector},
we obtain the following distributed space-time code matrix for the
proposed scheme:
\begin{equation}\label{eq:equivalent_S}
\bS=\left[ \bS_a, \cdots, \bS_a\right]
\end{equation}
where $\bS_a=\left[ \begin{array}{rr}
 s_1 & -s_2^*\\
 s_2 & s_1^*
\end{array}
\right]$ is the conventional Alamouti code matrix.

The channel  and relay power vectors are given by
\begin{eqnarray}\label{eq:halaK}
 \bh \!\!\!\!&=&\!\!\!\!\! [ f_1g_1, f_2^*g_2,
\ldots,f_{2K-1}g_{2K-1}, f_{2K}^*g_{2K}]^T\\
\label{eq:power-ala}
\bp \!\!\!\!&=&\!\!\!\! \left[p_1, p_2, \ldots , p_{2K-1},
p_{2K}\right]^T= \textstyle
\left[\sqrt{\frac{P_0P_1T}{m_{f_1}P_0+1}} b_1 \theta_1 ,\ldots ,
\sqrt{\frac{P_0P_{2K}T}{m_{f_{2K}} P_0+1}} b_K \theta_K \right]^T
\end{eqnarray}
respectively.

Note that in contrast to \eqref{eq:power_vector}, any $(2k-1)$th
and $(2k)$th relays use the same $b_k\theta_k$.

Conjugating the second entry $x_2$ of the vector $\bx=[x_1,x_2]^T$ in \eqref{eq:in-out}
and using \eqref{eq:equivalent_S}-\eqref{eq:power-ala},
we obtain the following equivalent model
\begin{equation}\label{eq:in-out2}
 \breve{\bx}=\bH\breve{\bs}+\breve{\bw}
\end{equation}
 where
$\breve{\bx}\triangleq [x_1,x_2^*]^T$, $\breve{\bs} \triangleq
[s_1,s_2^*]^T$, $\breve{\bw}\triangleq [w_1,w_2^*]^T$,
\begin{equation}
 \bH=\sum_{k=1}^K\bH_k
 \nonumber
\ \ \ \ \mbox{and} \ \ \ \ \
\bH_k=\left[ \begin{array}{cc}
 p_{2k-1} h_{2k-1} & -p_{2k} h_{2k}\\
 p_{2k}h_{2k}^* & p_{2k-1}h_{2k-1}^*
\end{array}\right].
\nonumber
\end{equation}

Note that
\begin{equation}
 \bH^H\bH=\left[ \begin{array}{cc}
 \gamma_a+\beta_a & 0\\
 0 & \gamma_a+\beta_a
\end{array}\right]
\nonumber
\end{equation}
where
\begin{eqnarray}\label{eq:alpha}
 \gamma_a \!\!\!\!&\triangleq&\!\!\!\! \|(\bp \odot \bh)\|^2\\
\label{eq:beta}
 \beta_a \!\!\!\!&\triangleq&\!\!\!\! \sum_{i,j=1,i\neq j}^{K}\theta_ib_i \theta_j b_j
\mathrm{Re}
 \{\rho_{2i-1,2j-1}h_{2i-1}h_{2j-1}^*
  +\rho_{2i,2j}h_{2i}h_{2j}^*\}.\quad\quad
\end{eqnarray}
Throughout (\ref{eq:equivalent_S})-(\ref{eq:beta}), the subindex
$(\cdot)_a$ stands for the extended Alamouti scheme.

Since the matrices $\tilde{\bA}_i$ satisfy the property
$\tilde{\bA}_i\tilde{\bA}_i^H=\bI_T$, the noise covariance matrix
$\bR_{\breve{\bw}} \triangleq \mbox{E}\{\breve{\bw}\breve{\bw}^H
\}$ is a scaled identity matrix. Therefore, the ML decoding
\begin{equation}
 \arg \min_{\breve{\bs}} \left \| \breve{\bx} - \bH\breve{\bs}
\right \| \nonumber
\end{equation}
reduces to simple symbol-by-symbol decoding.

As $|h_k|^2 = |\tilde{f}_k g_k|^2$, it is clear from
\eqref{eq:alpha}, \eqref{eq:beta} and Proposition 1 that the
maximum diversity can be achieved if $\beta_a \geq 0$. Similar to
\cite{AG04}, it can be proved that if $b_k \in \{-1,\,1\}$, it can
be guaranteed that $\beta_a \geq 0$.

As in Section \ref{sec:quasi}, the coefficients $b_k$ can be
selected using the exhaustive full search, a suboptimal SDR
approach similar to Algorithm 1, or an iterative procedure similar
to Algorithm 2. To develop the SDR approach for the extended
distributed Alamouti code case, we define the $K\times 1$ vector
$\bb_a \triangleq [b_1,\ldots,b_K]^T$ and the $2\times K$ matrix
\begin{equation}\label{eq:Fch}
    \bF\triangleq \left [ \begin{array}{cccc}
        p_1h_1 & p_3h_3 & \cdots & p_{2K-1}h_{2K-1} \\
        p_2h_2 & p_4h_4 & \cdots & p_{2K}h_{2K} \\
      \end{array}\right ].
\end{equation}
Using \eqref{eq:Fch}, we obtain that
\begin{equation}\label{eq:alpha2}
    \gamma_a+\beta_a = \bb_a^T \bF^H\bF \bb_a.
\end{equation}
Defining $\bar{\bQ}_a \triangleq \bF^H\bF$, we can write the
problem of optimal selection of the coefficients $b_k$
($k=1,\ldots,K$) as
\begin{equation}
    \max_{\bb_a \in \{-1,1\}^K} \bb^T_a \bar{\bQ}_a \bb_a.
    \nonumber
\end{equation}
Using the notation ${\bB}_a \triangleq \bb_a\bb^T_a$, this problem
can be rewritten as
\begin{eqnarray}\label{eq:op_alpha2}
    \max_{{\bB}_a} \,\trace({\bB}_a\bar{\bQ}_a)\quad
\mbox{s.t. }&\!\!\!\!\!\!\!\!& {\rm rank}\{ {\bB}_a \}= 1,\ \ {\bB}_a\succeq 0 \nonumber\\
    &\!\!\!\!\!\!\!\!&  {[{\bB}_a]}_{kk}=1, \ \  k=1,\,\ldots,\,K\
    \
\end{eqnarray}
and using the SDR approach, it can be approximately converted to a
convex form
\begin{eqnarray}\label{eq:op_alpha2new}
\max_{{\bB}_a} \,\trace({\bB}_a\bar{\bQ}_a)\  \mbox{s.t.
}&\!\!\!\!\!\!\!\!\!\!\!\!\!& {\bB}_a \succeq 0,\
{[{\bB}_a]}_{kk}=1,\, k=1,\,\ldots,\,K\ \ \quad
\end{eqnarray}
by omitting the rank-one constraint ${\rm rank}\{ {\bB}_a \}= 1$
in (\ref{eq:op_alpha2}).

The SDR-based algorithm for the proposed distributed Alamouti
approach be summarized as follows.

\underline{\textit{Algorithm 3}}
\begin{itemize}
\item[1.] At the receiver, find the solution to \eqref{eq:op_alpha2new}
using the approach of \cite{MDWLC02}.
\item[2.] Send the so-obtained $b_k$ from the receiver to the $k$th relay
pair for each $k=1,\ldots, K$ using one-bit per relay pair
feedback.
\end{itemize}

In turn, the greedy algorithm of Section \ref{sec:quasi} can be
modified as follows.

\underline{\textit{Algorithm 4}}
\begin{itemize}
\item[1.] Set $b_1=1$ and $\bm{\tau}_1=[h_1\, h_2]^T$.
\item[2.] For $k=2,\,\ldots,\,K$, compute
\begin{eqnarray*}
b_k= \mbox{sign}(\RE{[h_{2k-1}^*\,h_{2k}^*]
\bm{\tau}_{k-1}}),\ \ \
\bm{\tau}_k= \bm{\tau}_{k-1}+
b_k[h_{2k-1}\,h_{2k}]^T. \end{eqnarray*}
\item[3.] Send the so-obtained $b_k$ from the receiver to the $k$th relay
pair for each $k=1,\ldots, K$ using one-bit per relay pair
feedback.
\end{itemize}

To derive the {power loading coefficients} $\theta_k$, an approach
similar to that presented in Section \ref{sec:beam_weights} can be
applied. We first develop an approximation to the expected value
of the signal power and then maximize a lower bound on the SNR.
Using \eqref{eq:in-out2}, the average signal power can be written
as
\begin{equation}
\expv{P_{s}}=\expv{{\gamma}_a}+\expv{\beta_{a}}.
\end{equation}
Using \eqref{eq:alpha}, \eqref{eq:beta} and
the same arguments as in Section \ref{sec:beam_weights},
$\expv{P_{s}}$ can be approximated as
\begin{eqnarray}
\expv{P_s} \!\!\!\!&\approx&\!\!\!\!
 \sum_{i,j=1}^{K}\theta_i\theta_j|
\mathrm{Re}\{\expv{\rho_{2i-1,2j-1}h_{2i-1}h_{2j-1}^* +\rho_{2i,2j}h_{2i}h_{2j}^*}\}|\label{eq:Ps_approx_2}
\end{eqnarray}
where it is assumed that the optimal values of $b_i$
($i=1,\ldots,K$) are selected.

The expected value of the noise is given by
\begin{eqnarray}
\expv{P_w} \!\!\!\!&=&\!\!\!\! 1+\sum_{k=1}^{K}
 \theta_k^2 \left( \frac{P_{2k-1}m_{g_{2k-1}}}{P_0 m_{f_{2k-1}}+1} +
 \frac{ P_{2k}m_{g_{2k}}}{P_0 m_{f_{2k}}+1}
\right) .\ \ \ \ \ \label{eq:Pw_2}
\end{eqnarray}
Using \eqref{eq:Ps_approx_2} and \eqref{eq:Pw_2}, the SNR
maximization problem can be approximated as
\begin{eqnarray}\label{eq:op_SNRa}
    \max_{\bm{\theta}_a} \,\frac{\bm{\theta}_a^T\bQ_a \bm{\theta}_a}{\bm{\theta}_a^T\bW_a
    \bm{\theta}_a+1}\quad
\mbox{s.t. }\ \  \bar{\theta}_k\leq\theta_k^2\leq 1, \ i=1,\,\ldots,\,K
\end{eqnarray}
where $\bm{\theta}_a\triangleq [\theta_1\,\ldots\,\theta_K]^T$,
$$
\bW_a \triangleq \diag{\sum_{l=1}^2 \frac{m_{g_l} P_l
}{m_{f_l}P_0+1} ,\ldots, \sum_{l=2K-1}^{2K}\frac{m_{g_l}
P_l}{m_{f_l}P_0+1}}
$$
$\bQ_a$ is a $K\times K$ matrix with the entries
\begin{eqnarray}
[\bQ_a]_{i,j} \!\!\!\!&\triangleq&\!\!\!\!
\left | \RE{\rho_{2i-1,2j-1}\expv{h_{2i-1}h_{2j-1}^*}
+  \rho_{2i,2j}\expv{h_{2i}h_{2j}^*} } \right |
\label{eq:Qmata}
\end{eqnarray}
and $\bar{\theta}_k$ constrains the coefficients $\theta_k$ to
prevent diversity losses in a way similar to that described in
Section \ref{sec:quasi}. Now, the expected value in
\eqref{eq:Qmata} can be estimated using the statistical
independence of the channels as in \eqref{eq:exp_calc}. In
particular, for the $(2i,2j)$th factor in (\ref{eq:Qmata}), we
have
\begin{eqnarray}
\expv{h_{2i}h_{2j}^*}\!\!\!&=&\!\!\!
\left( \mu_{f_{2i}} \mu_{f_{2j}}^* + \delta_{(2i)(2j)}\sigma^2_{f_{2i}} \right)
 \left(
\mu_{g_{2i}}\mu_{g_{2j}}^*+\delta_{(2i)(2j)}\sigma^2_{g_{2i}}\right).\nonumber
\end{eqnarray}

Following the same steps as in Section \ref{sec:quasi}, the
optimization problem in \eqref{eq:op_SNRa} can be turned into a
convex feasibility problem that extends \eqref{eq:op_SNR3} to the
distributed Alamouti coding case.

\section{Differential transmission}\label{sec:differential}
The concept of differential transmission is used in this section
to extend the proposed approach to the case where no CSI is
available at the receiver. Let us assume that $T=1$ and let the
transmitter encode differentially the information symbols $s_l$
selected from some constant-modulo constellation $\mathcal{S}$ as
\begin{equation}
 u_l=u_{l-1}s_l, \quad u_0=1
 \nonumber
\end{equation}
where $u_l$ and $u_0$ are the current and initial transmitted
symbols, respectively. Similar to the coherent
scheme in \eqref{eq:in-out_bi}, we have
\begin{equation}\label{eq:diff1}
  x_l={\bf 1}^T_R(\bp \odot \bh) u_l + w_l.
\end{equation}
Using \eqref{eq:diff1}
and the previous received signal $x_{l-1}$, the ML
symbol estimate can be obtained from maximizing
\cite{LS03}
\begin{equation}
 \RE{x_{l-1}x_l^*s_l}
 \nonumber
\end{equation}
over $s_l \in \mathcal{S}$. We assume that no {power loading} is
used, i.e., set $\theta_i=1$ for $i=1,\ldots,R$. Since the
receiver has no CSI to select the feedback bits for $b_i$
($i=1,\ldots,R$), the following simple sequential feedback bit
assignment scheme can be used. Before the beginning of the frame
in which the information symbols should be transmitted, there is
an extra transmission stage to select the coefficients $b_i$.
First, $u_0$ is transmitted from the source to the relays and then
it is retransmitted by the relays to the destination with $b_i=1$
($i=1,\ldots, R$). Then, the second relay only alters its
coefficient $b_2$ to $-1$ and the relays retransmit again. The
received powers corresponding to the latter two
relay-to-destination transmissions are compared at the receiver
and the receiver sends one bit of feedback. This bit is used by
the second relay to select $b_2$ that corresponds to the maximum
received power. The process continues with the remaining relays in
the same way. This makes it possible to select all the
coefficients $b_i$ ($i=2,\ldots, R$) in a sequential (greedy) way.
After the process of selecting the coefficients $b_i$ is
completed, the source starts the transmission of its information
symbols. The overall transmission strategy can be summarized as
follows:

\underline{\textit{Algorithm 5}}
\begin{itemize}
\myitemsep
\item[1.] Set $b_i=1$, $i=1,\,\ldots,\,R$. Transmit $u_0$ from the
source to relays and then retransmit it from the relays to the
destination to obtain $x_1={\bf 1}^T_R(\bp \odot \bh) u_0 + w_1$
at the receiver.
\item[2.] For $j=2,\,\ldots,\, R$:
\vspace*{-1mm}
\begin{itemize}
 \myitemsep
\item[$\bullet$] At the $j$th relay, set $b_j=-1$ and, using (\ref{eq:ti}),
update the signal $d_j$ to be transmitted from this particular
relay.
\item[$\bullet$] Transmit signals from all relays to obtain
$x_j={\bf 1}^T_R(\bp \odot \bh) u_0 + w_j$ at the receiver.
\item[$\bullet$] If $|x_j|^2>|x_{j-1}|^2$, then feed ``$1$'' from
the receiver back to the relay; otherwise feed ``$0$'' back to the
relay. In the latter case, set $x_j=x_{j-1}$. \item[$\bullet$] If
the received feedback at the $j$th relay is $1$, then select
$b_j=-1$. Otherwise, select $b_j=1$.
\end{itemize}
\end{itemize}

Similarly, a differential modification of the extended distributed
Alamouti code of Section~\ref{sec:eAlamouti} can be developed in
the case when $T=2$. At the transmitter, a unitary matrix $\bS_l$
should be formed from the constant-modulo information symbols
$s_{2l-1},s_{2l}$ as
\begin{equation}
 \bS_l=\frac{1}{\sqrt{2}}\left[
\begin{array}{cc}
s_{2l-1} & -s_{2l}^*\\
s_{2l} & s_{2l-1}^*
\end{array}
\right]. \nonumber
\end{equation}

Let the differential encoding
\begin{equation}
 \bu_{l}=\bS_l\bu_{l-1}
 \nonumber
\end{equation}
be used at the transmitter. It amounts to sending the vector
$\bu_l=[u_{2l-1}, u_{2l}]^T$ instead of
 $\bs_l=[s_{2l-1}, s_{2l-1}]^T$ to the relays where $l$ denotes
the transmitted block number. The first vector $\bu_l$ can be
chosen as $\bu_0=[1, 0]^T$. Similar to \eqref{eq:in-out} and using
the matrices $\tilde{\bA}_{2k-1}$ and $\tilde{\bA}_{2k}$ defined
in the previous section for the extended distributed Alamouti
code, the following equivalent relation can be obtained
\begin{equation}
\bx_l=\bS_l \bU_{l-1} \left( \sum_{k=1}^K \bp_k\odot\bh_k\right) +
\bw_l \nonumber
\end{equation}
where $\bU_0=\bI_2$,
\begin{eqnarray*}\bU_{l-1}&\!\!\!\triangleq\!\!\!&\left[
\begin{array}{cc}
u_{2l-3} & -u_{2l-2}^*\\
u_{2l-2} & u_{2l-3}^*
\end{array}
\right], \quad l>1 \\
\bp_k &\!\!\!\triangleq\!\!\!& \left[ \sqrt{\frac{P_0 P_{2k-1}
T}{m_{f_{2k-1}} P_0+1}} b_k \theta_k,
 \sqrt{\frac{P_0 P_{2k} T}{m_{f_{2k}} P_0+1}} b_k \theta_k
\right]^T\\
\bh_k &\!\!\!\triangleq \!\!\!& [f_{2k-1}g_{2k-1},
f_{2k}^*g_{2k}]^T.
\end{eqnarray*}
The ML decoding amounts to maximizing \cite{LS03}
\begin{equation}
 \RE{ \trace\left(
\bx_{l-1}\bx_l^H \bS_l \right)} \nonumber
\end{equation}
over $s_{2l-1},s_{2l}\in \mathcal{S}$. Note that the detection can
be done symbol-by-symbol. As in the previous scheme without DSTC,
we set $\theta_i=1$ and use a similar strategy to select the
coefficients $b_i$ using relay pairs and blocks of length $T=2$.
This strategy can be summarized as follows:

\underline{\textit{Algorithm 6}}
\begin{itemize}
\myitemsep
\item[1.] Set $b_i=1$, $i=1,\,\ldots,\,K$. Transmit $\bu_0$ from the source to relays and
then retransmit it from the relays to the destination to obtain
$\bx_1=\bU_{0} \left( \sum_{k=1}^K \bp_k\odot\bh_k\right)  +
\bw_1$ at the receiver.
\item[2.] For $j=2,\,\ldots,\, K$:
\vspace*{-1mm}
\begin{itemize}
 \myitemsep
\item[$\bullet$] At the $(2j-1)$th and $(2j)$th relays, set $b_j=-1$ and, using
(\ref{eq:ti}), update the signals $\bd_{2j-1}$ and $\bd_{2j}$
 to be transmitted from this particular relay pair.
\item[$\bullet$] Transmit signals  from  all relays  to
obtain $\bx_j=\bU_{0} \left( \sum_{k=1}^K \bp_k\odot\bh_k\right) +
\bw_j$ at the receiver. \item[$\bullet$] If $\|\bx_j\|^2\geq
\|\bx_{j-1}\|^2$, then feed ``$1$'' from the receiver back to the
relay; otherwise feed ``$0$'' back to the relay. In the latter
case, set $\bx_j=\bx_{j-1}$. \item[$\bullet$] If the received
feedback at the $(2j-1)$th and $(2j)$th relays is $1$ then select
$b_j=-1$. Otherwise, select $b_j=1$.
\end{itemize}
\end{itemize}

Similar to Algorithms 2 and 4, Algorithms 5 and 6 are suboptimal.
However, the latter two algorithms do not require any CSI at the
receiver and, moreover, {our simulations demonstrate that they
achieve maximum diversity. It is also worth noting that this
diversity benefit is achieved at linear decoding complexity.}

\section{Simulations}\label{sec:simulations}
Throughout our simulation examples, the QPSK modulation is used
and the channels are assumed to be statistically independent from
each other. In all but the fourth example, we consider all the
channels to be complex circular Gaussian random variables with
zero-mean and unit variance and assume that $\theta_i=1$,
$i=1,\ldots,R$ (which are the optimal {power loading coefficients}
in this case). {For the sake of fairness of our comparisons, only
techniques that do not need the instantaneous CSI at the relays
are tested.} Unless specified otherwise, the feedback is
considered to be error-free.

\begin{figure}[t]
\begin{center}
  \includegraphics[width=8.5cm]{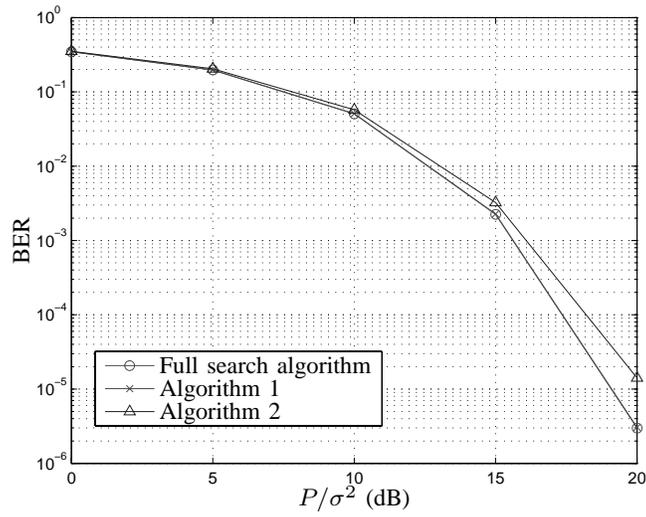}
\end{center}
\caption{BERs versus $P/\sigma^2$; first example.}
  \label{fig:Algorithms2}
\end{figure}

In the first example, we compare the bit error rate (BER)
performances of the algorithms that select the coefficients $b_i$
using the cooperative transmission scheme of
Section~\ref{sec:one_bit_feed} with $R=20$ relays and the same
maximum power $P_0=\ldots=P_R=P/(R+1)$. In this example, the full
search-based (optimal) algorithm is compared with Algorithms 1 and
2. Fig.~\ref{fig:Algorithms2} displays the BERs of these
algorithms $P/\sigma^2$. It can be seen from this figure that the
SDR-based approach (Algorithm 1) performs about 1 dB better than
the iterative procedure of Algorithm 2. The performances of the
optimal full search algorithm and Algorithm 1 are nearly
identical.

In our second example, the performances of the cooperative
transmission schemes of Sections~\ref{sec:one_bit_feed} (Algorithm
1) and \ref{sec:eAlamouti} (Algorithm 4) are compared with that of
the best relay selection (BRS) scheme, the distributed beamforming
approach of \cite{KJJ08} with quantized feedback, and the
distributed version of the QOSTBC \cite{JJ07}. {In the BRS scheme,
the destination selects the relay that enjoys the largest receive
SNR. The relays only have
 the knowledge of their average receive power $\expv{|r_i|^2}=
 m_{f_i}P_0+1$ and they use this knowledge to normalize the
 transmitted signal so that the average transmitted power of the $i$th relay
 is $P_i$. It can be readily shown that this corresponds to the
 following relay selection rule at the destination:
\begin{equation}
 \arg \max_{i=1,\ldots,R} \frac{|f_i g_i|^2 P_i}{1+
m_{f_i}P_0+|g_i|^2P_i}.
\end{equation}}

\begin{figure}[t]
\begin{center}
  \includegraphics[width=8.3cm]{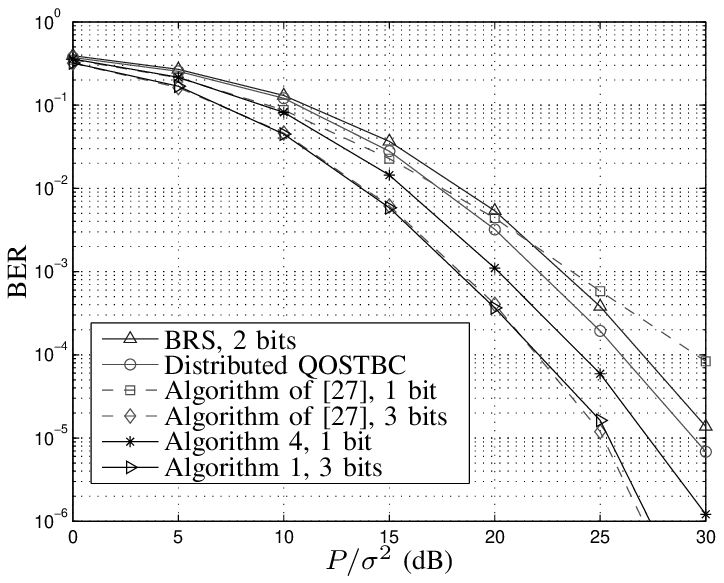}
\end{center}
\caption{BERs versus $P/\sigma^2$; second example.}
  \label{fig:PropvsSelec4a}
\end{figure}

Throughout this example, $R=4$ and the source and relay powers are
chosen from the optimal power distribution for DSTC \cite{JJ07} as
$P_0=P/2$ and $P_i=P/(2R)$ ($i=1,\ldots,R$). For the sake of
fairness, the distributed beamforming algorithm of \cite{KJJ08}
was implemented without the knowledge of the instantaneous channel
$f_i$ at each $i$th relay using the generalized Lloyd and genetic
algorithms. The beamformer codebooks required in the technique of
\cite{KJJ08} have been designed for the cases of one and three
feedback bits. Fig.~\ref{fig:PropvsSelec4a} displays the BERs of
the techniques tested versus $P/\sigma^2$.

\begin{figure}[t]
\begin{center}
  \includegraphics[width=7.5cm]{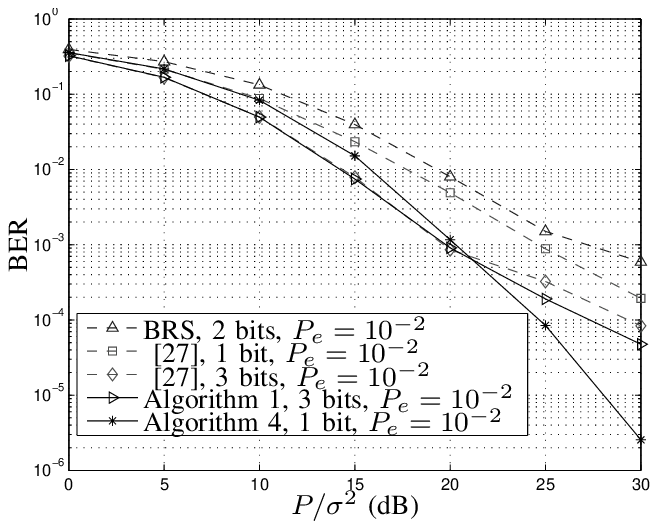}
\end{center}
 \caption{BERs versus $P/\sigma^2$; second example.}
  \label{fig:ErrorAnalysis2}
\end{figure}

{Note that the distributed QOSTBC technique does not require any
feedback, whereas the BRS technique requires two bits of feedback,
and the Algorithms 1 and 4 require three and one bits of feedback,
respectively. However, the distributed QOSTBC approach requires a
more complicated decoder and imposes the decoding delay of $T=4$.

It can be clearly seen from this figure that both Algorithms 1 and
4 substantially outperform BRS, distributed QOSTBC, and the
distributed beamforming approach of \cite{KJJ08} with one-bit
feedback. Also, Algorithm 1 outperforms Algorithm 4 with the
performance gain of more than 2 dB at the cost of a higher
feedback rate. The performances of Algorithm 1 and the approach of
\cite{KJJ08} with three bits of feedback are nearly identical.
However, it should be noted that the codebook design in the
technique of \cite{KJJ08} represents a rather difficult
optimization problem, and that this codebook has to be completely
redesigned and resent to the relay nodes whenever the channel
statistics or the transmitted powers change. This makes the
implementation of the beamformer of \cite{KJJ08} substantially
more difficult than that of our algorithms.

Fig.~\ref{fig:ErrorAnalysis2} compares the performance of our
Algorithms 1 and 4 with that of the BRS technique and the
beamformer of \cite{KJJ08} in the erroneous feedback case. All the
other parameters are the same as used in the previous figure. From
Fig.~\ref{fig:ErrorAnalysis2} we observe that our algorithms are
less sensitive to feedback errors than BRS and the approach of
\cite{KJJ08}.}

{In our third example, $R=4$ is chosen. In this example, the
performance of the differential techniques developed in Section
\ref{sec:differential} is compared to that of the Sp(2) DSTC of
\cite{JJ08d}, the coherent distributed QOSTBC of \cite{JJ07} and
the BRS technique with differential transmission in which the
relay with the largest received power is selected. Note that both
the Sp(2) DSTC and the coherent distributed QOSTBC do not require
any feedback, whereas the BRS approach requires a total of two
feedback bits. It should be also stressed that unlike our
differential schemes and the other schemes considered in this
example, the coherent distributed QOSTBC requires full CSI at the
receiver. The symbol rates of the Sp(2) DSTC, the coherent
distributed QOSTBC and the BRS technique are the same as that of
our differential techniques and are equal to $1/2$ symbols per
channel use. Important advantages of our technique w.r.t. the
Sp(2) DSTC are lower decoding complexity, shorter required channel
coherence time, and lower decoding delay.}

 For the Sp(2) code, we use
the 3-PSK constellation for the first two symbols and the 5-PSK
constellation for the other two symbols. With that, a total rate
of $0.9767$ bits per channel use (bpcu) is achieved. The other
schemes use the QPSK symbol constellations to achieve the total
rate of $1$ bpcu.

\begin{figure}[t]
\begin{center}
  \includegraphics[width=7.5cm]{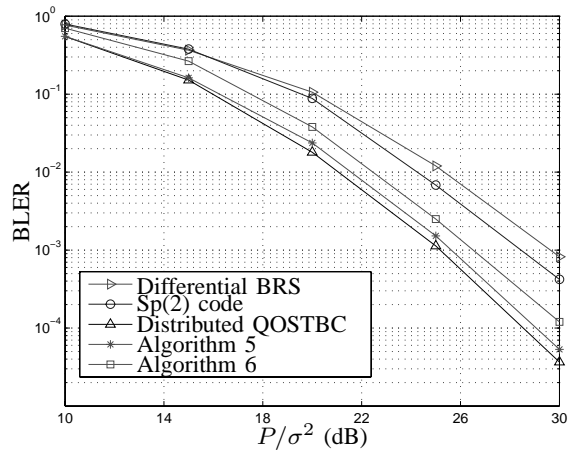}
\end{center}
  \caption{BLER versus $P/\sigma^2$; third example.}
  \label{fig:Differential2}
\end{figure}

Fig.~\ref{fig:Differential2} compares the block error rate (BLER)
performance of the techniques evaluated versus $P/\sigma^2$. The
values of BLER are computed using blocks of four symbols. As can
be observed from Fig.~\ref{fig:Differential2}, both Algorithms 5
and 6 outperform the Sp(2) code and the differential BRS approach,
and their performance is close to the distributed QOSTBC (which
requires the full CSI knowledge). {In particular, it can be seen
from this figure that the proposed techniques have approximately
the same (maximum) diversity order as the Sp(2) code and the
distributed QOSTBC with coherent decoder.}

 These improvements come
at the price of three bits and one bit of feedback for Algorithms
5 and 6, respectively. Also, note that Algorithm 5 uses a total of
$2R$ auxiliary time-slots before starting the transmission of
information bits, while Algorithm 6 uses $3K+1$ time-slots (one
time-slot for each feedback bit). On the other hand, the Sp(2)
code uses $2R$ auxiliary time-slots.

\begin{figure}[t]
\begin{center}
  \includegraphics[width=7.5cm]{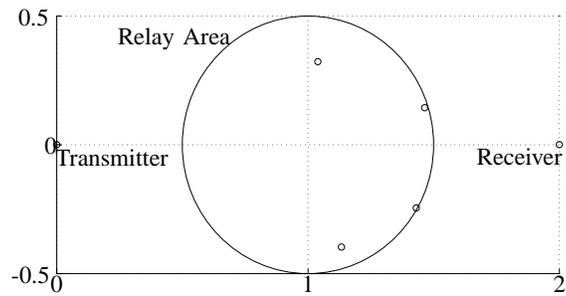}
\end{center}
  \caption{Geometry of the fourth example.}
  \label{fig:relay_loc2}
\end{figure}

In our fourth example, the performance of Algorithm 1 combined
with long-term {power loading} (which is developed in
Section~\ref{sec:beam_weights}) is compared with Algorithm 1 of
Section~\ref{sec:one_bit_feed} and with the analytical results
obtained from (\ref{eq:av_ser0}) by means of brute force
optimization. For long-term {power loading}, the approach of
(\ref{eq:op_SNR3}) with bisection search is used. In this example,
$R=4$ and $\bar{\theta}_1=\ldots=\bar{\theta}_R\triangleq
\bar\theta$, where the nearly optimal value of $\bar{\theta}=0.1$
has been chosen. The relay locations have been uniformly drawn
from a circle of normalized radius $0.5$, while the distance
between the source and destination is equal to $2$; see
Fig.~\ref{fig:relay_loc2} that explicitly clarifies the geometry.
The values of $m_{f_i}$ and $m_{g_i}$ depend on the distance from
the transmitter to the $i$th relay, where $m_{f_i}=m_{g_i}=1$ in
the center of the circle. We assume that the path-loss exponent is
equal to $3$. The performance is averaged over random channel
realizations whereas the relay locations are kept fixed over all
simulation runs. Both the line-of-sight (LOS) and non-LOS (NLOS)
scenarios are considered and equal maximum powers of the
transmitter and relay nodes ($P_0=P_1\ldots =P_R=P/(R+1)$) are
taken. In the LOS channel case, it is assumed that
$\phi_{f_i}=\phi_{g_i}=1$ where
$\phi_{f_i}\triangleq{|\mu_{f_i}|^2}/{\sigma_{f_i}^2}$ and
$\phi_{g_i}\triangleq {|\mu_{g_i}|^2}/{\sigma_{g_i}^2}$.

\begin{figure}[t]
\begin{center}
  \includegraphics[width=7.5cm]{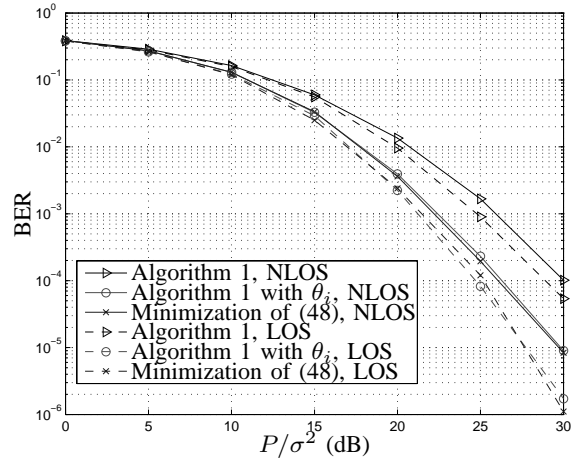}
\end{center}
  \caption{BER versus $P/\sigma^2$; fourth example.}
  \label{fig:OptvsDesign4a}
\end{figure}

In Fig.~\ref{fig:OptvsDesign4a}, the BERs of the algorithms tested
are shown versus $P/\sigma^2$. As can be clearly seen from the
figure, the proposed approach with long-term {power loading}
achieves nearly the same performance as predicted by
(\ref{eq:av_ser0}) and substantially outperforms Algorithm 1
without {power loading}.

In our fifth example, we compare the performances of Algorithm 1
of Section~\ref{sec:one_bit_feed} and Algorithm 4 of
Section~\ref{sec:eAlamouti} in the cases of perfect and imperfect
feedback. In this example, $R=4$, $P_0=P_1\ldots =P_R=P/(R+1)$ and
the feedback error probabilities $P_e=10^{-2}$ and $P_e=10^{-3}$
are assumed.

\begin{figure}[t]
\begin{center}
  \includegraphics[width=7.5cm]{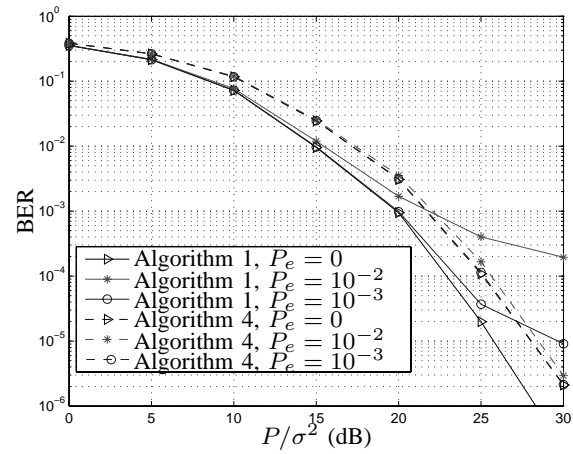}
\end{center}
  \caption{BER versus $P/\sigma^2$; fifth example.}
  \label{fig:ErrorAnalysis1}
\end{figure}

Fig.~\ref{fig:ErrorAnalysis1} displays the BERs of the methods
evaluated versus $P/\sigma^2$. As can be observed from this
figure, the performance of Algorithm 1 becomes sensitive to
feedback errors when the BER values are smaller than the feedback
error probability itself. Therefore, as the same link quality can
be normally expected in both directions, the performance of
Algorithm 1 should not be significantly affected by feedback
errors.

It can also be seen from Fig.~\ref{fig:ErrorAnalysis1} that, in
contrast to Algorithm 1, the performance of Algorithm 4 is not
sensitive to feedback errors. The latter fact can be explained by
the spatial diversity of the Alamouti code.

\section{Conclusions}\label{sec:conclusions}

A new approach to the use of a low-rate feedback in wireless relay
networks has been proposed. It has been shown that our scheme
achieves the maximum possible diversity offered by the relay
network. To further improve the performance of the proposed scheme
in practical scenarios, the knowledge of second-order channel
statistics has been used to obtain long-term {power loading}
coefficients by means of maximizing the receiver signal-to-noise
ratio with proper power constraints. This maximization
 problem has been shown to be approximately equivalent to
 a convex feasibility problem whose solution has been demonstrated
  to be close to the optimal one in terms of the error
probability. To improve the robustness of our scheme against
feedback errors and further decrease the feedback rate, an
extended version of the distributed Alamouti code has been
developed. Finally, extensions of the proposed approach to the differential
transmission case have been discussed.

Simulations have verified an improved performance-to-feedback
tradeoff of the proposed techniques as compared to other popular
techniques such as distributed QOSTBC of \cite{JJ07}, best relay
selection method, distributed beamforming technique of
\cite{KJJ08} with quantized feedback, and the Sp(2) distributed
code of \cite{JJ08d}.

\section*{Appendix\\
Proof of Proposition 1}
The symbol error probability (SER) for
\eqref{eq:in-out_bi} is given by \cite{SA05}
\begin{equation}
 \mathrm{SER}=c_1 Q\left( \sqrt{c_2 \frac{P_s}{P_w}}\right)
 \nonumber
\end{equation}
where $c_1$ and $c_2$ are two constants that depend on the
constellation used, and $Q(x)=\frac{1}{2\pi}\int_{x}^{\infty}
e^{-t^2/2} dt$.

Using the Chernoff bound, we have
\begin{equation}\label{eq:av_ser0}
 \overline{\mathrm{SER}}\leq \frac{c_1}{2}
\expvv{e^{-c_2\frac{P_s}{2 P_w}}}{f_i,g_i}.
\end{equation}

Note that if we establish an upper bound for $\beta=0$, then it
will be also valid for any $\beta>0$. This follows from the fact
that $Q(x)$ is a monotonically decreasing function. Using this
fact, let us obtain an upper bound on $\overline{\mathrm{SER}}$ by
using the particular value $\beta=0$. Then, from
(\ref{eq:av_ser0}) we obtain
\begin{equation}\label{eq:av_ser}
 \overline{\mathrm{SER}}\leq \frac{c_1}{2}
\expvv{e^{-c_2\frac{\gamma}{2 P_w}}}{f_i,g_i}.
\end{equation}

First, let us calculate the expected value over the channel
coefficients $f_i$. As these coefficients are statistically
independent, each term in the sum for $\gamma$ can be calculated
independently. Using the complex Gaussian pdf for $f_i$
\begin{equation}
 p_{f_i}(f_i)=\frac{1}{\pi \sigma_{f_i}^2}
e^{-|f_i-\mu_{f_i}|^2/\sigma_{f_i}^2}
\end{equation}
and defining
\begin{equation} \label{eq:a_i}
 a_i \triangleq \frac{c_2 |g_i|^2 \rho_{i,i}}{2P_w}
\end{equation}
we obtain from (\ref{eq:av_ser}) that
\begin{equation} \label{eq:expcv_g}
 \overline{\mathrm{SER}}\leq \frac{c_1}{2} {\rm E}_{g_i}\!\left\{\prod_{i=1}^{R}
\Upsilon_i\right\}
\end{equation}
where
\begin{equation}\label{eq:Upsilon}
 \Upsilon_i \triangleq \frac{1}{\pi \sigma_{f_i}^2}\int_{-\infty}^{\infty}
e^{-a_i |f_i|^2 -|f_i-\mu_{f_i}|^2/\sigma_{f_i}^2}\, df_i.
\end{equation}
After straightforward manipulations, (\ref{eq:Upsilon}) can be
rewritten as
\begin{eqnarray}\label{eq:Upsilon1}
\Upsilon_i &\!\!\!\!=\!\!\!\!& \frac{1}{a_i \sigma_{f_i}^2+1}
e^{-\frac{|\mu_{f_i}|^2}{\sigma_{f_i}^2} \left(
\frac{a_i\sigma_{f_i}^{2}}{a_i \sigma_{f_i}^{2}+1}\right)}
\int_{-\infty}^{\infty}\!\!
\frac{a_i\sigma_{f_i}^2+1}{\pi\sigma_{f_i}^2}\,
e^{-\frac{(a_i\sigma_{f_i}^2+1)}{\sigma_{f_i}^2}\left
|f_i-\frac{\mu_{f_i}}{(a_i\sigma_{f_i}^2+1)}\right |^2} df_i.
\end{eqnarray}
The function inside the integral in (\ref{eq:Upsilon1}) is equal
to the complex Gaussian pdf
$\mathcal{CN}\left(\frac{\mu_{f_i}}{a_i\sigma_{f_i}^2+1},
\frac{\sigma_{f_i}^2}{a_i\sigma_{f_i}^2+1}\right)$. Therefore, the
integral in (\ref{eq:Upsilon1}) is equal to one and we obtain that
\begin{equation}
 \Upsilon_i = \frac{1}{a_i \sigma_{f_i}^2+1}
e^{-\phi_{f_i} \left( \frac{a_i\sigma_{f_i}^{2}}{a_i
\sigma_{f_i}^{2}+1}\right)} \nonumber
\end{equation}
where $\phi_{f_i}={|\mu_{f_i}|^2}/{\sigma_{f_i}^2}$. An upper
bound approximation for the expected value in \eqref{eq:expcv_g}
can be derived as follows. Since $a_i\geq 0$, we have that $
 0 \leq \frac{a_i\sigma_{f_i}^{2}}{a_i \sigma_{f_i}^{2}+1} < 1
$. Therefore, $\Upsilon_i$ can be upper-bounded as $\Upsilon_i\le
1/(a_i \sigma_{f_i}^2+1)$ and
\begin{equation}
 \overline{\mathrm{SER}}\leq \frac{c_1}{2}{\rm E}_{g_i}\!\left\{\prod_{i=1}^{R}
\frac{1}{a_i \sigma_{f_i}^2+1} \right\}.\label{SERoverl}
\end{equation}
Let us characterize the power of each transmitting node $P_i$
($i=0,\ldots,R$) as a fraction $P_i= \lambda_i P$ of the total
power $P=\sum^{R}_{i=0} P_i$, where $\sum_{i=0}^R \lambda_i=1$. If
$R$ is large, then, according to the law of large numbers,
\begin{equation}\displaystyle \sum_{i=1}^{R} \frac{\lambda_i
|g_i|^2}{m_{f_i}\lambda_0+P^{-1}} \leq \,\, R \alpha \nonumber
\end{equation} where the inequality is
satisfied in the almost sure sense,
$$\displaystyle\alpha \triangleq \max_{i=1,\ldots,R} \left(
\frac{\lambda_i m_{g_i}}{m_{f_i}\lambda_0+P^{-1}} \right) $$ and
$m_{g_i}\triangleq \expv{|g_i|^2}=|\mu_{g_i}|^2+\sigma_{g_i}^2$.
Therefore, from \eqref{eq:noise_pow1} and \eqref{eq:a_i}, we have
that
\begin{eqnarray}
 \frac{1}{{a_i\sigma_{f_i}^2+1}}=
\frac{1}{\frac{c_2 \sigma_{f_i}^2 |g_i|^2 \rho_{i,i}}{2P_w} +
1}\leq \frac{1}{\frac{c_2 \sigma_{f_i}^2 |g_i|^2
\rho_{i,i}}{2(1+R\alpha)}+1}. \label{ineq}
\end{eqnarray}
Using (\ref{rho}) and (\ref{ineq}), from (\ref{SERoverl}) we
obtain that
\begin{equation}\label{eq:ser_g}
 \overline{\mathrm{SER}}\leq\frac{c_1}{2}{\rm E}_{g_i}\!\left\{\prod_{i=1}^{R}
\frac{1}{\bar{a}_i P |g_i|^2 +1} \right\}
\end{equation}
where
$$ \displaystyle
\bar{a}_i \triangleq \frac{c_2 \sigma_{f_i}^2 \lambda_0
\lambda_i}{ 2(m_{f_i}\lambda_0+1/P)(1+R\alpha)}.
$$

{From} \eqref{eq:ser_g} and the fact that all the channel
coefficients are statistically independent, it can be readily seen
that for each $i$ the expectation over $g_i$ in the right-hand
side can be calculated independently from the other values $g_l$,
$l\ne i$. The random variable $z_i=|g_i|^2$ has the non-central
chi-square pdf with two degrees of freedom:
\begin{equation}
 p_{|g_i|^2}(z_i) = \frac{1}{\sigma_{g_i}^2}
e^{-\frac{z_i+|\mu_{g_i}|^2}{\sigma_{g_i}^2}} I_0\!\left(
\frac{2|\mu_{g_i}|\sqrt{z_i}}{\sigma_{g_i}^2}\right)
\label{pdf_chi}
\end{equation}
where $I_0(\cdot)$ denotes the modified zero-order Bessel function
of the first kind. Using (\ref{pdf_chi}) to compute the
expectation in \eqref{eq:ser_g}, we have
\begin{equation}
 \overline{\mathrm{SER}}\leq\frac{c_1}{2}\prod_{i=1}^{R} \bar{\Upsilon}_i
 \nonumber
\end{equation}
where
\begin{eqnarray}
 \bar{\Upsilon}_i &\!\!\!\triangleq\!\!\!&
\int_{0}^{\infty}\frac{1}{\bar{a}_i P z_i +1}\, p_{|g_i|^2}(z_i)\,
dz_i= e^{-\phi_{g_i}}\int_{0}^{\infty} \frac{e^{-y_i}}{ (\bar{a}_i
P \sigma_{g_i}^2 y_i +1)} I_0\!\left( 2
\sqrt{\phi_{g_i}y_i}\right) dy_i \label{eq:int2}
\end{eqnarray}
$y_i\triangleq z_i/\sigma_{g_i}^2$ and $\phi_{g_i}\triangleq
{|\mu_{g_i}|^2}/{\sigma_{g_i}^2}$.  Let us break up the integral
\eqref{eq:int2} into two terms as
$\int_{0}^{\infty}=\int_{0}^{1/P} +\int_{1/P}^{\infty}$ and use
the following results from \cite{JJ08c} to approximate
$\bar{\Upsilon}_i$:
\begin{eqnarray}
\int_{0}^{1/P} \!\!e^{-y_i}I_0\left( 2 \sqrt{\phi_{g_i}y_i}\right)
dy_i= \frac{1}{P}+\mathcal{O}(P^{-2})\approx
\frac{1}{P}\label{eq:integral1}
\\
\int_{1/P}^{\infty}\!\! y_i^{-1}e^{-y_i} I_0\left( 2
\sqrt{\phi_{g_i}y_i}\right) \!dy_i\!=\!
  E_1(P^{-1})+\sum_{k=1}^{\infty}\frac{\phi_{g_i}^k}{k! k}\,\,\,\,\,\label{eq:integral2}
\end{eqnarray}
where $E_1(q)\triangleq
-\tilde{c}-\log{q}-\sum_{k=1}^{\infty}\frac{(-1)^{k}q^k}{k! k}$
for $q>0$,
 and $\tilde{c}$ denotes the Euler's constant. Note that if $\log{P}\gg
1$, then $E_1(P^{-1})\approx \log{P}$. Using the latter fact,
{from} \eqref{eq:int2}-\eqref{eq:integral2} we obtain for the case
of large $P$ that
\begin{equation}\label{eq:ser_end2}
 \overline{\mathrm{SER}}\leq\frac{c_1}{2}\prod_{i=1}^{R}
\frac{e^{-\phi_{g_i}}}{\bar{a}_i  \sigma_{g_i}^2 }
\frac{(\log{P}+q_i)}{P}
\end{equation}
where $q_i\triangleq \bar{a}_i\sigma_{g_i}^2 +
\sum_{k=1}^{\infty}\frac{\phi_{g_i}^k}{k! k}$. Defining
$$\kappa\triangleq \frac{c_1}{2}\prod_{i=1}^{R}
\frac{e^{-\phi_{g_i}}}{\bar{a}_i  \sigma_{g_i}^2 }$$ and using the
properties of the logarithm, we can rewrite \eqref{eq:ser_end2} as
\begin{equation}
\label{finaleq}
 \overline{\mathrm{SER}}\leq \kappa\left(
P^{-R(1-\frac{\log{\log{P}}}{\log{P}})}+\left(
\prod_{i=1}^{R}q_i\right)P^{-R}\right).
\end{equation}
For large values of $P$, the first term in the sum in
(\ref{finaleq}) will dominate. Hence, Proposition 1 is proved.
\hfill $\blacksquare$

\bibliographystyle{IEEEbib}

\end{document}